\documentclass[prl,twocolumn,showpacs,floatfix,superscriptaddress,showpacs]{revtex4}

\usepackage{graphicx}
\usepackage{epsfig}

\begin{document}
\newcommand{\norm}[1]{\ensuremath{| #1 |}}
\newcommand{\aver}[1]{\ensuremath{\langle #1 \rangle}}
\newcommand{\ket}[1]{\ensuremath{| #1 \rangle}}
\newcommand{\bra}[1]{\ensuremath{\langle #1 |}}
\newcommand{\e}{\ensuremath{\textrm{e}}}
\renewcommand{\i}{\ensuremath{\textrm{i}}}
\title{Scanning tunnelling microscopy for ultracold atoms}
\author{Corinna Kollath}
\affiliation{DPMC-MaNEP, University of Geneva, 24 Quai
Ernest-Ansermet, CH-1211 Geneva, Switzerland}
\author{Michael K\"ohl}
\affiliation{Cavendish Laboratory, University of Cambridge, JJ
Thomson Avenue, Cambridge CB3 0HE, United Kingdom}
\affiliation{Institute of Quantum Electronics, ETH Z{\"u}rich,
CH-8093 Z{\"u}rich, Switzerland}
\author{Thierry Giamarchi}
\affiliation{DPMC-MaNEP, University of Geneva, 24 Quai
Ernest-Ansermet, CH-1211 Geneva, Switzerland}
\date{\today}

\begin{abstract}

We propose a novel experimental probe for cold atomic gases
analogous to the scanning tunnelling microscope (STM) in condensed
matter. This probe uses the coherent coupling of a single particle
to the system. Depending on the measurement sequence, our probe
allows to either obtain the \emph{local} density, with a
resolution on the nanometer scale, or the single particle
correlation function in real time. We discuss applications of this
scheme to the various possible phases for a two dimensional
Hubbard system of fermions in an optical lattice.
\end{abstract}
 \maketitle

%paragraph 1

Recent advances in the field of ultracold atoms have led to a
close connection between quantum gases and condensed matter
physics. The achievement of strongly correlated systems and their
remarkable tunability open the possibility to realize `quantum
simulators' for quantum many-body phenomena. To name one example,
ultracold fermionic systems clarified the crossover between a
BCS-state of paired fermions to a Bose-Einstein condensate of
ultracold bosonic molecules
\cite{RegalJin2004,BartensteinGrimm2004,ZwierleinKetterle2004,KinastThomas2004,BourdelSalomon2004}.
Further investigations of strongly correlated systems were
initialized by the successful loading of ultracold bosonic
\cite{GreinerBloch2002} and fermionic \cite{KoehlEsslinger2005}
atoms into three-dimensional optical lattices. In these periodic
lattice potentials created by counter-propagating laser beams the
physics of different lattice models can be mimicked
\cite{JakschZoller1998,HofstetterLukin2002}. In particular the
fermionic Hubbard model, which plays an important role on the way
of understanding high-temperature superconductivity, can be
realized %quite
naturally given the short range nature of the
interactions between the neutral atoms.

%paragraph 2
Whereas achieving the exotic quantum phases experimentally appears
feasible with today's technology their clear identification
remains an obstacle. Compared to condensed matter the neutrality
of the cold atoms is both an advantage and a drawback since they
cannot be perturbed as easily as electrons in a solid. Possible
probes are thus more sophisticated than their condensed matter
counterparts. In addition, for ultracold atomic gases an
inhomogeneous confining potential causes the coexistence of
different spatially separated quantum phases
\cite{JakschZoller1998,FoellingBloch2006,CampbellKetterle2006}.
This makes their realization and observation in the presence of a
trapping potential very involved and creates a need for a method of
probing the systems locally. However, the existing probes
\cite{FoellingBloch2006,CampbellKetterle2006} still involve an
averaging over regions of various densities and new techniques
which allow for a local detection need to be developed.

%paragraph3
In condensed matter physics, a remarkable local probe was provided
by the scanning tunneling microscope (STM)
\cite{BinnigRohrer1982}. It allowed to explore and image the
surface topography with atomic resolution, paving the way to
control and analyze quantum phenomena on solid surfaces
\cite{CrommieEigler1993}. In addition to the density analysis with
unprecedented resolution, the STM has also become a spectroscopic
tool probing the local density of states. This spectroscopic
method had a major impact on the understanding of the physical
properties of strongly correlated systems for which the local
density of states provides unique information on the physics of the
system. In particular the STM has made significant contributions
to the field of high temperature superconductors
\cite{FischerBerthod2006}.

%paragraph 4

In this work we propose a novel experimental setup to locally
probe cold atomic systems in an approach similar to and as
versatile as the STM. The probe relies on the coupling of a single
particle to the system. Different `operating modes' yield either a
measurement of the local density or of the single particle Green's
function in time. The realization of such a probe will open the
possibilities to investigate exotic quantum phases in great detail
as we show on the example of the Hubbard model. In extension to
the conventional STM in condensed matter physics our scheme would
allow for measurements in a three-dimensional sample.

%paragraph 5

\begin{figure}
\begin{center}
{\includegraphics[width=0.75\columnwidth]{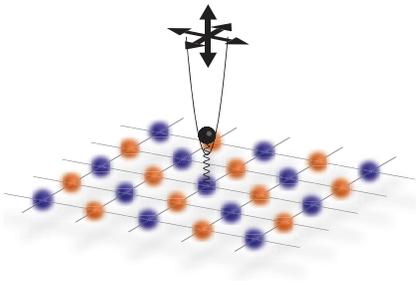}}
\end{center}
\caption{Sketch of the cold atom tunneling microscope. As an example the
  application to an anti-ferromagnetic state with alternating spin states
  labeled by different colors is shown.} \label{fig:STM}
\end{figure}

The key idea for the realization of an STM-like scheme with cold
atoms, the `cold atom tunneling microscope' is sketched in
Fig.~\ref{fig:STM}. A single trapped particle is used as a probe
of local quantities by inducing a controlled interaction between
the probe particle and the quantum many-body state. To allow for a
precise control over the motion of the probe and to facilitate a
convenient readout mechanism, we suggest to employ a single atomic
ion trapped in the vibrational ground state of a radio-frequency
Paul trap~\cite{WinelandMeekhof1998}. In this case the spatial
resolution of the microscope relies on the excellent control over
the position and motion of trapped ions on the sub-micron scale.
However, the working principle of the microscope does not depend
on the charge of the probe particle and it also applies to a
neutral atomic quantum dot \cite{RecatiZoller2005} provided that
the trapping potential of the dot has only a negligible influence
on the quantum many-body system \cite{BrudererJaksch2006}. The
controlled interaction between the probe particle (the ion) and
the quantum many-body system could be provided by a two-photon
Raman coupling.

%paragraph 6

As we show below the cold atom tunnelling microscope facilitates a
local detection of the density on individual lattice sites and,
quite remarkably, it even allows to perform a spin-resolved
detection of the density. The realization of a spin-resolved STM
is a long-sought goal in condensed matter systems but has not yet
been achieved. The cold atom tunnelling microscope also allows to
perform spectroscopy by observing the local single particle
Green's function $\aver{c^\dagger_{\sigma,j}(t_0)
c_{\sigma,j}(0)}_F$ in time. Here $c_{\sigma,j}$ is the
annihilation operator for the neutral atom on a site $j$ with spin
$\sigma=\{\uparrow, \downarrow\}$ and $\aver{\cdot}_F$ stands for
taking the expectation value with respect to the atomic system
only. The temporal decay of this function directly reflects the
nature of the excitations and gives thus direct information on the
quantum phases present in the system.

\begin{figure}
\begin{center}
{\epsfig{figure=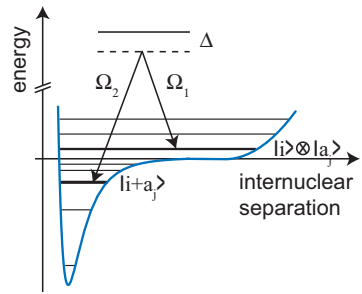,width=0.6\columnwidth}}
\end{center}
\caption{Two-photon Raman coupling of the ion $|i\rangle$ and an
atom $|a_j\rangle$ in a harmonic potential well to a molecular ion
bound state $\ket{i+a_j}$. The single photon coupling is detuned
by $\Delta$ from a resonant transition to suppress spontaneous
emission from the intermediate excited state. The effective
two-photon Rabi frequency $\Omega_0$ is proportional to the
coefficients of the single photon transitions and inversely
proportional to the detuning $\Delta$, i.e.~$\Omega_0 \propto
\Omega_1 \Omega_2/\Delta$.} \label{fig:sequencescanning}
\end{figure}

%paragraph 7
%`Scanning mode': Measuring the local density
We first show taking the example of fermionic atoms in two
different spin states $\uparrow$ and $\downarrow$ in an optical
lattice how a measurement of the local density, the 'scanning mode', can be achieved. It
is facilitated by a two-photon Raman coupling between the ion
$\ket{i}$ and an atom $\ket{a_j}$ in a lattice well $j$ by which a
weakly bound molecular ion $|i+a_j\rangle$ can be created (see
Fig.~\ref{fig:sequencescanning}). This coupling can be described by the expression
$\left( \sum_\sigma \Omega_\sigma(t) M_\sigma^\dagger I
c_{\sigma,j}+h.c.\right)$. Here $M_\sigma$ and $I$ are the
annihilation operators for the molecular ion and the atomic ion,
respectively. The coupling strength $\Omega_\sigma(t)$ can be
controlled experimentally. By choosing the correct frequency and
polarization of the laser fields, the coupling is dependent on the
atomic `spin' state paving the way for the spin-resolved
microscopy.

%paragraph 8

The experimental sequence to detect the local density is as
follows: At time $t=0$ the atomic many-body system is prepared in
its ground state $\ket{\Psi_0}$. The ion is introduced into the
lattice well $j$ in state $\ket{i}$ and the Raman coupling is
switched on for a duration $\delta t$,
i.e.~$\Omega_{\sigma}(t)=\Omega_{\sigma,0}$ if $t\in [0,\delta t]$
and vanishes otherwise. The time $\delta t$ has to be short
compared to the internal time-scales of the probed system (i.e.
time-scales set by the atom-atom interaction $U$, the kinetic
energy of the atoms $J$, and the atom-ion contact interaction
$U_{ai}$) to avoid a change of the many-body state during the
probe sequence.  The Raman coupling generates a superposition of
the initial state $\ket{\Psi_0}\otimes\ket{i}$ and the state
$c_{\sigma,j}\ket{\Psi_0}\otimes\ket{i+a_j}$ in which one atom is
removed from the system and a molecular ion is formed. The ratio
between the amplitudes of the two states depends on the density of
atoms in the well $j$. Hence detecting the probability (i.e. the
average of the outcome of several quantum measurements) for
molecule formation after the application of the Raman pulse
measures the local density of atoms in the lattice well $j$ by the
relation $\aver{\sum_\sigma M_\sigma^\dagger M_\sigma}=
\sum_\sigma \sin^2{(\Omega_{\sigma,0} \delta t)}
\aver{n_{\sigma,j}}_F$ with $\aver{n_{\sigma,j}}$ the local atomic
density. The outcome of the photo-association process can be
detected by measuring the changed oscillation frequency of the
heavier molecular ion in the Paul trap or by observing the absence
of resonant light scattering of the molecular ion and its
reappearance after photo-dissociation \cite{SugiyamaYoda1997}. The
procedure can be repeated scanning different lattice sites as
sketched in Fig.~\ref{fig:STM} with a spatial resolution on the
order of 20\,nm \cite{Eschner2001}. To facilitate the measurement
of the density with a good signal to noise ratio, the lattice
potential could be increased  such that the density profile on
different lattice sites is frozen and sequential measurements on
single lattice site are feasible.

\begin{figure}
\begin{center}
{\epsfig{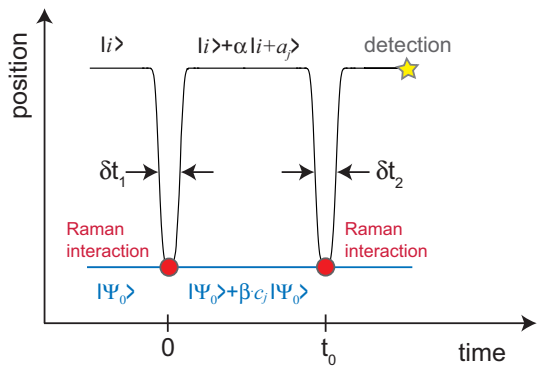}}
\end{center}
\caption{Schematics of the experimental sequence for the tunnelling
mode. The atomic ion $|i\rangle$ is introduced at the lattice site
$j$ into the many-body system in state $|\Psi_0\rangle$. The
two-photon Raman process (red) couples an atom at this lattice
site $|a_j\rangle$ to the ion with a certain amplitude.
Subsequently, the ion and the many-body system are separated for
the probe time $t_0$ during which they evolve individually. After
recombination, the Raman interaction is applied again and the
molecule formation is detected.} \label{fig:sequencetunneling}
\end{figure}

%paragraph 9
%\section{`Tunneling mode': Probing the local single particle Green's function}
Using the cold atom tunnelling microscope with a different
sequence, the 'tunneling mode', allows to perform spectroscopy and
to measure time dependent correlations locally. The experimental
sequence is sketched in Fig.~\ref{fig:sequencetunneling}. As in
the scanning mode we start at $t=0$ in the state
$\ket{\Psi_0}\otimes \ket{i}$, i.e. the ground state
$\ket{\Psi_0}$ of the atomic system and a single atomic ion. A
two-photon Raman process is applied over a short time interval
$\delta t_1$ to couple the ion with an atom present in the lattice
well. Subsequently, the superposition state of the atomic and the
molecular ion $|i\rangle +\alpha |i+a_j\rangle$ is removed from
the system such that they are non-interacting with the remaining
quantum many-body system, for example their center-of-mass
position can be shifted by applying a small dc voltage. After a
variable time of free evolution $t_0$ in this isolated position
they return into the addressed lattice well and the application of
the two-photon Raman process is repeated for a time interval
$\delta t_2$. The outcome of the molecule formation is detected
afterwards \footnote{Note that most of the terms do only appear if
the system under consideration is a quantum many body system.}:
\begin{widetext}
\begin{eqnarray}
\label{eq:full}
\hspace{-0.5cm} \aver{M^\dagger M} &&= A(\delta t_1,\delta t_2) + \sin^2(\delta t_2 \Omega_{0})\left[\cos (\delta t_1 \Omega_{0})-1\right]\left\{[\cos (\delta t_1 \Omega_{0})-1] \aver{n_{j}(0)n_{j}(t_0)n_{j}(0)} +\aver{n_{j}(t_0)n_{j}(0)}\right\}\nonumber \\
 &&+\left[\sin^2(\delta t_2 \Omega_{0}) +\sin^2(\delta t_1 \Omega_{0}) \right]\aver{n_{j}(t_0)}+\sin^2(\delta t_1 \Omega_{0})\left[\cos(\delta t_2
\Omega_{0})-1\right]\aver{  c^\dagger_{j}(0) c_{j}(t_0)
  c^\dagger_{j}(t_0) c_{j}(0) } \\
\hspace{-0.5cm} A(\delta t_1,\delta t_2)&&=2\sin(\delta t_1 \Omega_{0})\sin(\delta t_2 \Omega_{0})\cos(\delta t_2\Omega_{0})\Re \left \{
    \e^{-i(\varepsilon_M-\varepsilon_I)t_0/\hbar} \right. \left.
\left[ \underbrace{\aver{c^\dagger_{j}(t_0) c_{j}(0)}}_{=:A_1}
\right.\right.
\left.\left. +(\cos(\delta t_1 \Omega_{0})-1) \underbrace{\aver{n_{j}(0)
    c^\dagger_{j}(t_0) c_{j}(0)}}_{=:A_2}\right]\right\} \nonumber
\end{eqnarray}
\end{widetext}
We supressed the spin index and on the right hand side additionally the index $F$.

%paragraph 10

Using appropriate measurement sequences different correlation functions can be extracted. To obtain the temporal correlation function
$\aver{c^\dagger_{\sigma,j}(t_0) c_{\sigma,j}(0)}_F$ the described
measurement procedure is applied sequentially: first, using
$\delta t_1=\delta t_2=\delta t$ for both pulses, second using
$\delta t_1=\delta t$ for the first pulse and $\delta
t_2=2\pi/\Omega_{\sigma,0}-\delta t$ for the second pulse. Subtracting
the outcome for the molecule formation of the two measurements gives
$\Delta \aver{M_\sigma^\dagger M_\sigma} = 2 A(\delta t,\delta t)$. For small values of $(\delta t \Omega_{\sigma,0})$ the pre-factor
of the first summand $A_1$ in $A(\delta t,\delta t)$ is quadratic in $(\delta t
\Omega_{\sigma,0})$, whereas the prefactor of the second term $A_2$ is quartic. Since
additionally in many systems the decay of the correlation function
$\aver{n_{\sigma,j}(0) c^\dagger_{\sigma,j}(t_0) c_{\sigma,j}(0)}$
is faster or comparable to the decay of the single particle
correlation function the second term can safely be neglected.

%paragraph 11

The expression $(\varepsilon_I-\varepsilon_M)t_0/\hbar$ represents
the phase difference the atomic ion and the molecular ion collect
during the time $t_0$. In principle this quantity could be zeroed
by choosing a suitable combination of the optical lattice field
and the ion trapping fields. However, this cancellation is not
necessary if $\varepsilon_I-\varepsilon_M \gg U,J$ because then
the temporal evolution of the correlation function is encoded
simply in the envelope of the detection signal.

%paragraph 12
%Application to correlated quantum phases
One direct application of the cold atom tunnelling microscope would
be the identification of the quantum many-body phases of the two
dimensional Hubbard model. In
addition to the normal (Fermi liquid) quantum fluid of fermions,
this model can lead to broken symmetry phases such as an
anti-ferromagnet, and a strongly correlated (Mott) insulator. An
important and yet open question is whether other more exotic
phases can exist in this model, such as inhomogeneous distribution
of the density (stripes and checkerboards) or even superconducting
phases with d-wave symmetry for the pairing. Our local probe directly detects symmetry broken
phases such as the anti-ferromagnet in which the spin density is
modulated (cf.~Fig.~\ref{fig:STM}) and even more inhomogeneous
phases with a modulation of the density (stripes and checkerboards
\cite{HoffmanDavis2002}).

\begin{figure}
\begin{center}
{\psfig{figure=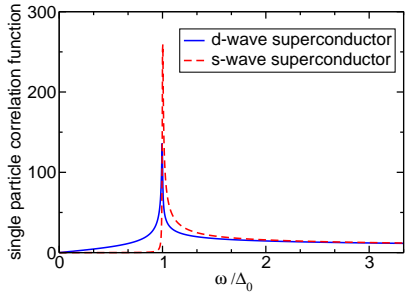,width=0.53\columnwidth,angle=-90}}
\end{center}
\caption{The Fourier transform of the temporal correlation
function in an $s$-wave and a $d$-wave superconducting state is
shown for a gap value of $\Delta_0=0.3J$. } \label{fig:supercond}
\end{figure}

%paragraph 13
Additionally, even for phases with homogeneous density and spin
density, such as a quantum fluid or a superconductor, the
'tunnelling mode' reveals the nature of the excitations by probing
the single particle density of states. This, for example, allows
to characterize directly an s-wave or a d-wave superconductor. In
Fig.~\ref{fig:supercond} we plot the Fourier transform of the
correlation function $\aver{c^\dagger_{\sigma,j}(t_0)
c_{\sigma,j}(0)}_F$ for both an s-wave superconducting and a
d-wave superconducting phase on a two-dimensional lattice. Both
are obtained using the phenomenological BCS-approach using the
energy dispersion on the lattice $-2J(\cos(k_xa)+\cos(k_y a))$. In
the s-wave superconducting phase, the gap $\Delta_s(k)\equiv
\Delta_0$, clearly leads to a strong divergence of the correlation
signal and a zero response in the gap  below $\Delta_0$. For the
d-wave superconducting phase with $\Delta_d(k)=\frac{\Delta_0}{2}
(\cos(k_xa)-\cos(k_y a))$ \cite{SigristUeda1991}, one observes a
quite different signal having a spectral weight below the gap
energy. Thereby the structure of the superconducting order
parameter and the size of the gap can be extracted from the
proposed measurement.

%paragraph 14
%Experimental implementation
The independent control over the single particle and the neutral
atomic quantum gas lies at the heart of the cold atom tunnelling
microscope. To a good approximation the ion experiences only the
ion trapping potential, the atom only the optical lattice
potential and the weakly bound molecular ion both potentials. This
makes the single ion a particularly attractive choice for
shuttling the atomic and the molecular ion in and out of the
lattice without influencing the neutral atomic quantum many-body
state. For example a displacement of 1.2\,mm within 50\,$\mu$s has
been achieved without exciting vibrational quanta \cite{Rowe2002}.

%paragraph 15
The binding energies of the weakly bound states of the atom-ion
interaction potential (see Fig.~\ref{fig:sequencescanning}) are
determined by its asymptotic behavior scaling as $-C_4/r^4$. Here
$C_4$ is proportional to the electric dipole polarizability of the
neutral atom and $r$ is the internuclear separation. The binding
energy of the most weakly bound molecular state is two orders of
magnitude less than for typical neutral atom interactions
\cite{GribakinFlambaum1993,CoteLukin2002}. Several more deeply
bound states with binding energies in the 10-100\,MHz range are
available for Raman photo-association \cite{LeRoyBernstein1970}.
The generation of weakly bound molecules using two-photon Raman
coupling in optical lattices has already been demonstrated for
pairs of neutral atoms \cite{RomBloch2004} and even the coherent
coupling of free atomic and bound molecular states has been
observed \cite{RyuHeinzen2005} which is the prerequisite for the
tunneling mode.

%paragraph 16
In order to probe the quantum many-body state without
perturbations the time scales set by the parameters of the atomic
system should be larger than the time intervals of the Raman
pulses. To realize a strongly correlated phase in the lattice, the
atom-atom scattering length $a_{aa}$ needs to be enhanced by a
Feshbach resonance. Assuming $a_{aa} \approx a_{ai} \approx
10^3$\,$a_0$ results in $U \simeq U_{ai} \simeq 20$\,kHz for the
fermionic isotope $^{40}$K, whereas $J$ is typically one order of
magnitude smaller. Therefore the condition for the proposed
`scanning' mode $J$, $U$, $U_{ai}\,\ll 1/\delta t$ can for example
be fulfilled using an effective Raman coupling $\Omega_{\sigma,0}
= 2 \pi \times 10\,$kHz applied over a time interval $\delta t =
5\,\mu$s resulting in a molecule formation probability of $\approx
0.1$. For the `tunnelling' mode the Raman coupling needs to be one
order of magnitude stronger since the above condition has to be
fulfilled for both Raman pulses of duration $\delta t$ and
$2\pi/\Omega_{\sigma,0}-\delta t$, respectively. The shortness of
the photo-association pulse has other direct benefits: first, the
level shift due to the interaction $U_{ai}$ is not resolved and
thus the measurement is independent of the occupation of the
lattice well by an atom in the second hyperfine state which is not
probed in the spin-resolved mode. Secondly, the short pulse and
the subsequent removal of the molecular ion from the quantum
many-body system ensures also the stability of the microscope
scheme against three-body recombination in a lattice well.

%paragraph 17
In conclusion, we have proposed a novel experimental setup, the
cold atom tunnelling microscope, to observe \emph{locally} the
(spin-resolved) density and the single particle Green's function.
In contrast to previous work this measurement procedure does not
average over spatially different regions of the system with
coexisting quantum phases, but can resolve single lattice wells. A
modification of the proposed scheme would give also access to
\emph{nonlocal} single particle correlation functions. The
required modification consists of moving the probe particle during
the tunnelling mode scheme to a different lattice well, say $m$,
before the second Raman pulse is applied. The outcome of the
molecule formation then will be related to the correlation
function $\aver{c^\dagger_{\sigma,m}(t_0) c_{\sigma,j}(0)}_F$ of
the atomic system. Additionally the proposed setup opens the
possibility to create single particle excitations in a controlled
way.

\acknowledgments

We would like to thank C.~Berthod and H. H\"affner for fruitful
discussions. This work was partly supported by the SNF under MaNEP
and Division II.

%\bibliography{references}

\end{document}